# Atomic and electronic structures of stable linear carbon chains on Ag-nanoparticles

D.W. Boukhvalov[1,2], I.S. Zhidkov[3], E.Z. Kurmaev[3,4], E. Fazio[5], S.O. Cholakh[3], and L. D'Urso[5]

[1]*Department of Chemistry, Hanyang University, 17 Haengdang-dong, Seongdong-gu, Seoul 04763, Korea*
[2]*Theoretical Physics and Applied Mathematics Department, Ural Federal University, Mira Street 19, 620002 Yekaterinburg, Russia*
[3]*Institute of Physics and Technology, Ural Federal University, Mira Street 19, 620002 Yekaterinburg, Russia*
[4]*M.N. Mikheev Institute of Metal Physics, Russian Academy of Sciences, Ural Branch, S. Kovalevskaya Street 18, 620990 Yekaterinburg, Russia*
[5]*Dipartimento di Scienze Matematiche e Informatiche, Scienze Fisiche e Scienze della Terra (MIFT), Università di Messina, 98166 Messina, Italy*
[6]*Dipartimento di Scienze Chimiche, Universita` di Catania, Viale Andrea Doria 6, 95125 Catania, Italy*

*In this work, we report X-ray photoelectron (XPS) and valence band (VB) spectroscopy measurements of surfactant-free silver nanoparticles and silver/linear carbon chains (Ag@LCC) structures prepared by pulse laser ablation (PLA) in water. Our measurements demonstrate significant oxidation only on the surfaces of the silver nanoparticles with many covalent carbon-silver bonds but only negligible traces of carbon-oxygen bonds. Theoretical modeling also provides evidence of the formation of robust carbon-silver bonds between linear carbon chains and pure and partially oxidized silver surfaces. A comparison of theoretical and experimental electronic structures also provides evidence of the presence of non-oxidized linear carbon chains on silver surfaces. To evaluate the chemical stability, we investigated the energetics of the physical adsorption of oxidative species (water and oxygen) and found that this adsorption is much preferrable on oxidized or pristine silver surfaces than the adsorption of linear carbon chains, which makes the initial step in the oxidation of LCC energetically unfavorable.*

E-mail: danil@hanyang.ac.kr

# 1. Introduction

Low-dimensional materials are a focus in current material science research. [1,2] Linear one-dimensional carbon chains (LCCs) are the next step after the development of two-dimensional carbon systems like graphene and its derivatives. [2] Recent theoretical works have discussed the unusual physical properties of LCCs. [3] One of the methods for the production of LCCs is to etch a graphene monolayer until a monoatomic chain is formed between etched areas. [4] This approach requires state-of-the-art processing techniques and is rather difficult to reproduce and transfer from laboratory to industry. The successful synthesis of graphene monolayers (see for review [5]), other 2D materials [6], and novel compounds such as silicene [7] and germanene [8] on metallic substrates has motivated researchers to use these approaches for the synthesis of LCCs.

Previous workers showed that silver nanoparticles have unique properties for some reactions, their surface robustly interacts with a great variety of species, which both, prevent aggregation and oxidation of nanoparticles [9] and catalyze the synthesis of low-dimensional structures, like carbon-based nanomaterials. This approach has provided additional motivation for the attention towards carbon-coated metals and might represent a new class of materials that could be employed for nanotechnology applications. This aspect is also important in biological systems [10], for example, encapsulated carbon-silver/gold nanospheres are interesting candidate for SERS-active tag for ultrasensitive detection of biomolecules and in vivo imaging [11]

Silver/linear carbon chain (core/shell) structures can be prepared by pulsed laser ablation (PLA) in water. This technique offers an eco-friendly way to synthesize surfactant-free colloids. Raman characterization results and atmospheric pressure matrix-assisted laser desorption ionization MS experiments (AP-MALDI) for such systems demonstrated that LCCs contain *sp*-hybridization either as alternating triple and single bonds (polyynes) or with consecutive double bonds (cumulenes). [12,13] However, these characterizations do not provide detailed information about the atomic and electronic structure of LCCs on silver surfaces ~~substrates~~, which is required for a complete understanding of their physical and chemical properties. The nature of carbon-metal interactions also remains unclear. Previous experimental [14,15] and theoretical [15,16] studies on carbon-based materials, demonstrated an enhancement in the chemical activity of graphene in the presence of a metallic substrate and especially in areas with defects [15]. Theoretical works predicted the instability of free-standing LCCs in an oxidative environment [17,18], which was further confirmed by experiments. [19] Surprisingly, reported in Ref. [11] LCCs on silver surfaces are stable in water. In this last case, a few nanometer layer of carbon species (sp and sp2 hybridized carbon) was detected on the surface of silver nanoparticles, sometimes as a bridge between nanoparticles [12,13]. It is reported that, in the absence of a metal end-capping stabilization, LCCs tend to react and crosslink toward $sp^2$ hybridizations in about 2h. [20] Instead, in LCC-Me hybrid nanostructures the presence of silver-carbon interactions hind their degradation, therefore resulting in stable systems for several months [21]. Understanding the nature of this chemical stability will both satisfy scientific curiosity and be very helpful for increasing the chemical stability of low-dimensional systems. [22]

In the present study, we present the results of XPS measurements (core levels and valence bands) of Ag and Ag@LCC nanoparticles prepared by pulsed laser ablation in water. Obtained results are compared with density functional theory (DFT) calculations of Ag@LCC structures and modeling of the adsorption of oxygen and water on carbon atoms and on silver in the vicinity of LCC.

**2. Experimental and Calculation Details**

Surfactant-free silver nanoparticles (NPs) and silver/linear carbon chain LCC (core/shell) structures were prepared using nanosecond pulsed laser Nd:YAG 532 nm (operating at a repetition rate of 10 Hz with a pulse width of 5 ns) ablation (PLA) in liquid water. For Ag@LCCs preparation, the same metal colloidal dispersion was used as a medium to ablate a polycrystalline graphite (99.99% of purity) target (for details of the preparation see. Ref. [12]).

The morphology of the samples was investigated by scanning transmission electron microscopy (STEM). STEM analyses were carried out with a Zeiss-Gemini 2 electron microscope, operating at 150 kV. Few drops of an aqueous suspension, containing the prepared nanostructures, were deposited on nickel grid to perform morphology characterization.

XPS core-level and valence band (VB) measurements were performed using a PHI 5000 Versa Probe XPS spectrometer (ULVAC Physical Electronics, USA) based on a classic X-ray optic scheme with a hemispherical quartz monochromator and an energy analyzer working in the binding energy range of 0 to 1500 eV. This apparatus uses electrostatic focusing and magnetic screening to achieve an energy resolution of $\Delta E \leq 0.5$ eV for Al K$\alpha$ radiation (1486.6 eV). The analytical chamber was pumped down using an ion pump to achieve a pressure lower than $10^{-7}$ Pa. The XPS spectra were recorded using Al K$\alpha$ X-ray emission. The spot size was 200 µm, and the X-ray power load delivered to the sample was less than 50 W. Typical signal-to-noise ratios were greater than 10000:3. A few drops from the nanoparticle suspension was deposited and dried on a monocrystalline silicon substrate, to record X-ray photoelectron spectra (XPS) of the core-levels and valence bands. The presence of silver-carbon bonds is of crucial importance in the solid state analysis since free sp chains air-exposed easily degrade as a consequence of their interaction with atmospheric gases [20]. Thus, we are confident that the obtained LCC species on silver nanoparticles are stable and suitable for XPS analysis.

The modeling was carried out using density functional theory embedded in the pseudopotential code SIESTA. [22] All calculations were performed using the GGA-PBE approximation [23] while considering van der Waals corrections. [24] All calculations were performed in spin-polarized mode, and the energy mesh cut-off was 360 Ry in a 4×4×2 k-point mesh using the Monkhorst–Pack scheme. [25] During optimization, the electronic ground state was found to be self-consistent using norm-conserving pseudopotentials [26] for cores; a double-ζ plus polarization basis of localized orbitals for silver, carbon, and oxygen; and a double-ζ basis for hydrogen. Optimization of bond lengths and total energies were performed with an accuracy of 0.04 eV/Å and 1 meV, respectively.

For the modeling of large nanoparticles, we used slabs of 4 layers within periodic boundary conditions. The number of layers corresponds to the minimal number of internal layers, which is greater than the number of surface layers. The separation distance between the slabs is governed by six lattice parameters that have values on the order of 2 nm. In the simulation of a free-standing LCC, we used a supercell with periodic boundary conditions along the x-axis. Chains were separated by 2 nm of empty space. The binding energy is defined by the standard formula:

$$E_{bind} = [(E_{Ag(001)} + E_{LCC}) - E_{Ag@LCC}]/N,$$

where $E_{Ag(001)}$ is the energy of a pristine or oxidized silver (001) surface, $E_{LCC}$ is the energy of a free-standing LCC, $E_{Ag@LCC}$ is the energy of an Ag@LCC system, and N is the number of carbon atoms in an LCC in the supercell. Note that positive values of binding energies indicate stable configurations.

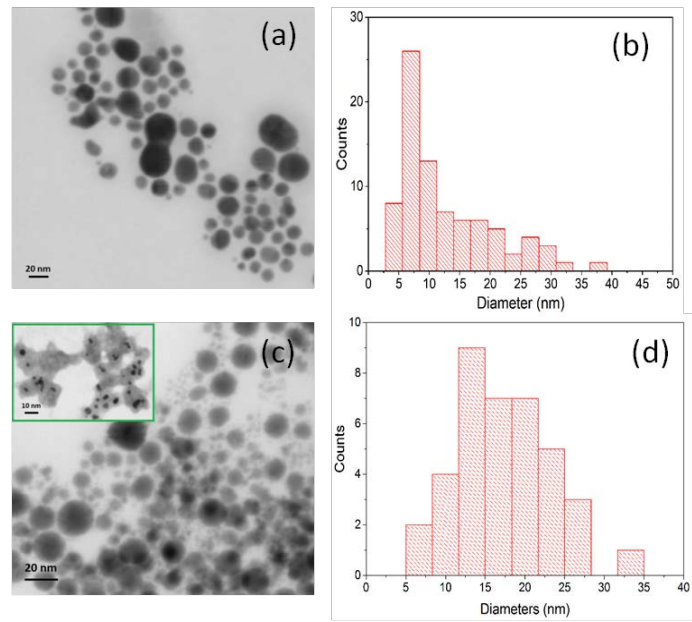

**Figure 1.** STEM micrograph of as-prepared Ag (a) and Ag@LCC (c) nanostructures and the histograms reporting the obtained size distribution

## 3. Results and Discussion

Figure 1(a) indicates that as prepared STEM picture of Ag Nps is spherical, well separated, with a dimension less than 40 nm and an average particle size of about 10-15 nm. When carbon is ablated in the presence of silver colloids, nanoparticles appeared reduced in size (Fig 1(c)), with an average particles diameter of about 8 nm, organized in ensembles of aggregates with interparticle distances of a few nanometers (see inset in figure 1(c)). We attributed the nanoparticle size reduction to the laser-induced fragmentation during the second irradiation process; in these conditions, it is possible that the formation of ensembles can be ascribed to aggregation phenomena mediated by the presence of carbon species interacting with the surface of Ag Nps. Also, we do not exclude that during the second irradiation a series of Ag end-capped LCCs are also formed together to the Ag Nps encapsulated in carbonaceous shells.

XPS measurements of survey spectra (Fig. 2) show the presence of C, O, Si, Cu, Ag, and N signals. The surface compositions determined from these spectra are presented in Table I.

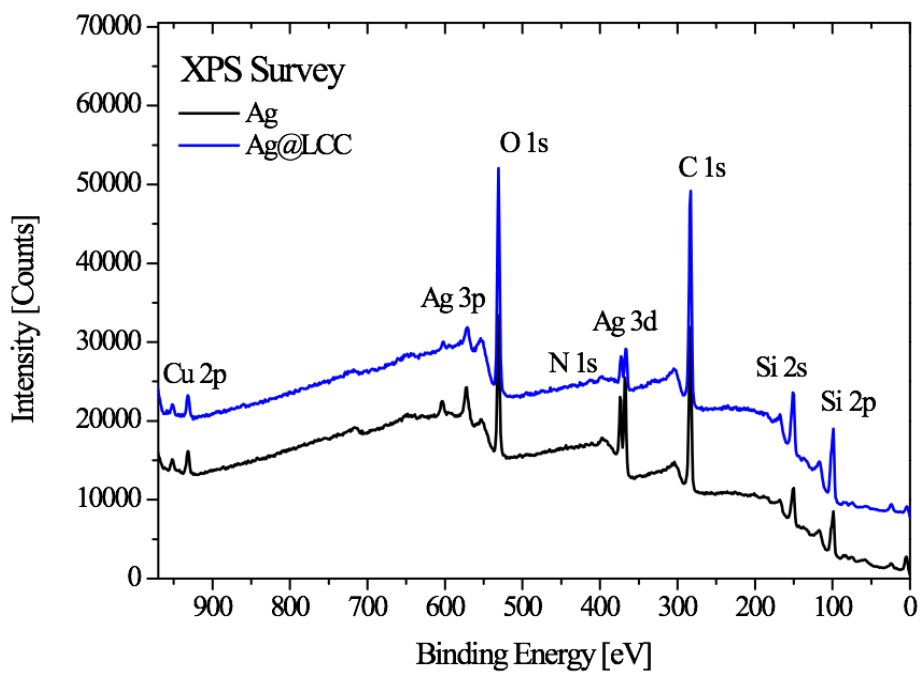

**Figure 2.** XPS survey spectra of Ag and Ag@LCC nanoparticles.

**Table I.** Surface composition of Ag and Ag@LCC nanoparticles (in at%).

| Sample | C | O | Si | Cu | Ag | N |
|--------|------|------|------|-----|-----|-----|
| Ag | 61.2 | 21.4 | 14.0 | 0.9 | 2.4 | - |
| Ag@LCC | 56.2 | 24.6 | 16.7 | 0.6 | 0.8 | 1.0 |

High-energy resolved XPS Ag 3d spectra of Ag and Ag@LCC nanoparticles are presented in Fig. 3. The XPS Ag $3d_{5/2}$ and Ag $3d_{3/2}$ core level binding energies for Ag NPs appeared at 368.1 and 374.1 eV, respectively, which is in good agreement with bulk silver metallic values [27]. On the other hand, XPS spectra of Ag@LCC NPs showed a second set of Ag $3d_{5/2,3/2}$ lines with binding energies of 366.5 and 372.4 eV. The appearance of additional lines cannot be attributed to oxidation of Ag@LCC nanoparticles because XPS Ag 3d lines in oxides have higher binding energies in XPS spectra, e.g., Ag $3d_{5/2}$=367.33 eV for AgO [28] and 367.62 eV for $Ag_2O$ [29].

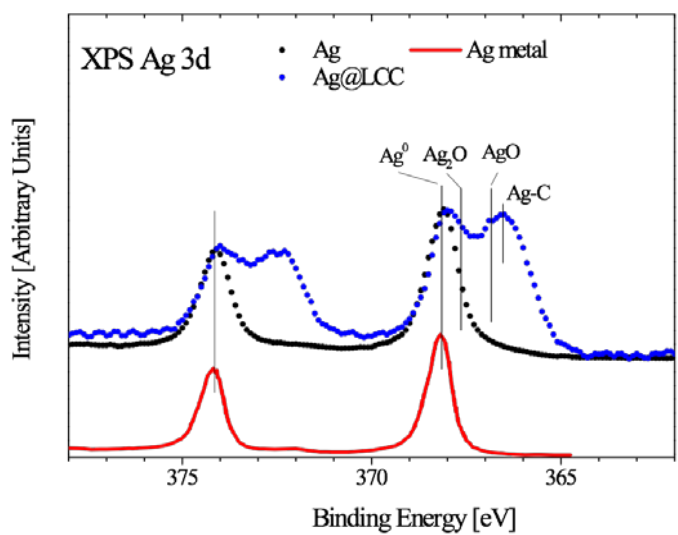

**Figure 3.** XPS Ag 3d-spectra of Ag and Ag@LCC nanoparticles.

This conclusion is supported by the XPS O 1s measurements (Fig. 4), which are found to be almost identical for Ag and Ag@LCC nanoparticles.

The peak position at 531.8 eV corresponds to [OH]-related species [30] due to water adsorption on the surfaces of Ag-and Ag@LCC nanoparticles. As seen from Fig. 4, the contributions of AgO and $Ag_2O$ in the XPS O 1s and C-O (C=O) are negligible. Therefore, there was no oxidation of Ag or Ag@LCC nanoparticles or formation of C-O and C=O bonds.

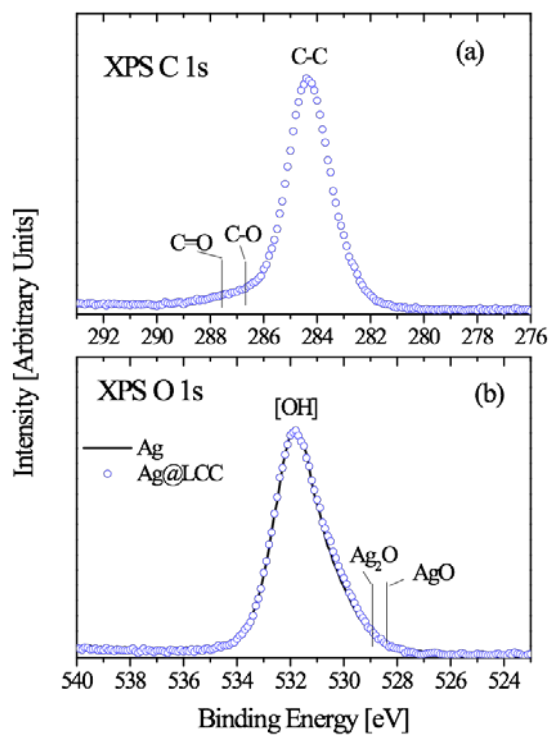

**Figure 4.** XPS O 1s of Ag and Ag@LCC nanoparticles.

In our opinion, the appearance of additional peaks of XPS Ag $3d_{5/2,3/2}$ lines of Ag@LCC NPs can be attributed to the formation of Ag-C bonds. Formation of Ag-C bonds in Ag nanoparticles was also found in previous studies [31-33]. Kawai et al. [31] measured XPS Ag 3d-spectra of water-dispersible silver nanoparticles stabilized by metal-carbon σ-bonds and found binding energies of Ag $3d_{5/2,3/2}$ lines at 366 and 372 eV, which correlates well with our measurements. The formation of Ag-C bonds in Ag@LCC NPs was also confirmed by XPS valence band measurements (Fig. 5). The main features of XPS VBs of Ag NPs were very similar to those of metallic silver [34], whereas the XPS VB of Ag@LCC NPs was strongly modified by a low-energy shift of the highest peak. This is probably due to hybridization of Ag 4d and C 2p-states and the appearance of a high-energy C 2s sub-band. It is interesting to note that the energy difference of these features (7.9 eV) is close to the energy separation of C 2p and C 2s orbital energies (ΔE=8.5 eV) in free carbon atoms [35], which also agrees with our conclusion about the formation of Ag-C bonds in Ag@LCC nanoparticles.

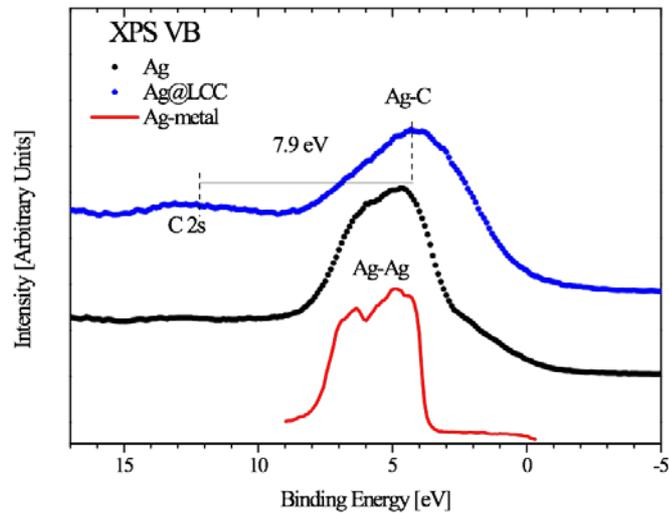

**Figure 5.** XPS VB spectra of Ag and Ag@LCC nanoparticles.

**Table II.** Calculated carbon-silver distances and carbon-carbon distances within chains (in Å), binding energy between LCC and substrate (in meV/carbon atom, note positive values correspond to favorable binding), change of Mulliken populations (in e$^-$) on carbon atoms and difference (in meV/molecule) in total energy between physisorption over an area near an LCC (Fig. 6c) and over carbon atoms (Fig. 6b). Negative values correspond to a favorable first configuration.

| Substrate | d(C-Ag) | d(C-C) | $E_{bind}$ | $\Delta e^-$ | $\Delta E_{O2}(\Delta E_{H2O})$ |
|---|---|---|---|---|---|
| Pristine (Fig. 6a) | 2.42 | 1.361 | 264 | +0.026 | -616 (-1250) |
| Passivated by oxygen (Fig. 6b) | 2.57 | 1.375 | 137 | -0.088 | -376 (-94) |
| Passivated by hydroxyl groups (Fig. 6c) | 2.23 | 1.372 | 414 | +0.023 | -533 (-1503) |

Experimental measurements provide important but insufficient information about the local atomic structure of LCCs and the nature of their chemical stability. The first step of our theoretical analysis was to compare the electronic structures of modeled Ag@LCC nanoparticles with experimental results. We assessed three possible configurations: LCC over pristine (001) silver surfaces (Fig. 6a) and the same surface passivated by oxygen atoms (Fig. 6b) or by hydroxyl groups (Fig. 6c). The choice of models is based on Ag 3d spectra (Fig. 3), which demonstrate the formation of multiple Ag-O bonds.

The calculated distances between carbon and silver atoms and the binding energies (Table II) demonstrate the formation of robust bonds between LCC and metallic nanoparticles, which agrees with the Ag 3d spectra (Fig. 3). Formation of this bonds provides increasing of carbon-carbon distances in LCC from 1.293 Å in a freestanding chant to the values of order 1.37Å. The value of increasing the bond length is the same order as previously reported alteration of bond length in phenyl-terminated polyenes on silver substrates. [36] Note that the binding energy per carbon atom is higher than in the case of graphene on metallic substrates [37] The magnitude of binding energy is proportional to the magnitude charge transfer between substrate and LCC (Tab II). On the other hand, XPS spectra of graphene sheet/metal interfaces show the absence of metal-carbon bonds. [38] Comparison of the values of transferred electrons[36,37] demonstrate the decrease of the magnitude of charge transfer with the grown of the size carbon system on the metallic substrate that provides weakening of carbon-metal bonding.

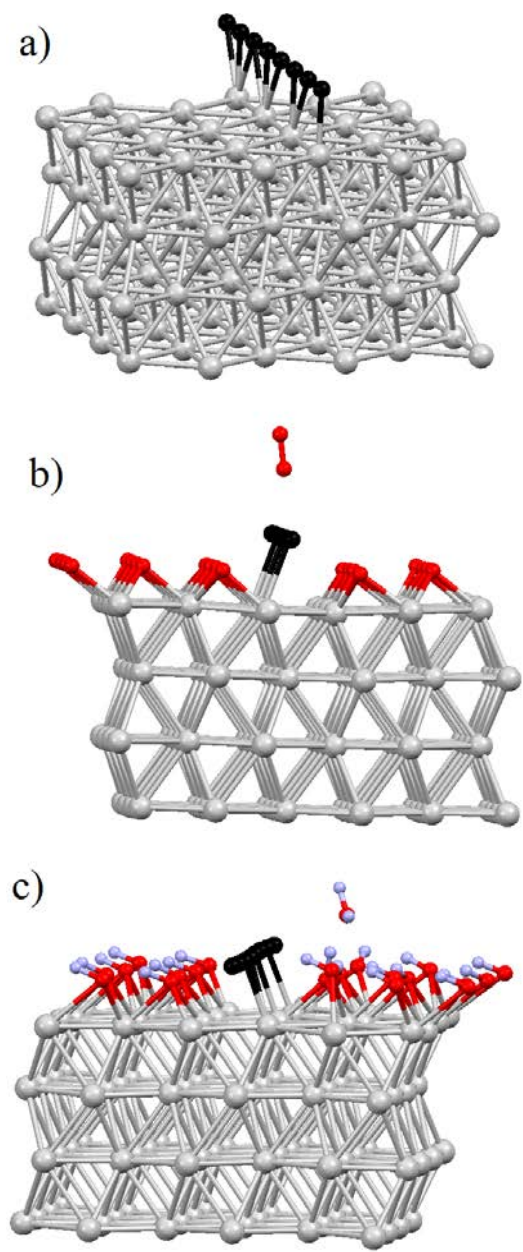

**Figure 6.** (a) Optimized atomic structure of a linear carbon chain over pristine silver, (b) the same LCC on a silver surface passivated by oxygen with oxygen molecules with a physisorbed LCC and that (c) LCC on a silver surface passivated by hydroxyl groups with water molecules physisorbed over the area.

The last step of our modeling was to understand the nature of the chemical stability of Ag@LCC. The first step of any oxidation process is physical adsorption of the oxidative species over the site of further oxidation. We compared two possible sites for the adsorption of oxidative species: the first is over a carbon atom (Fig. 6b), and the second is in the vicinity of the LCC (Fig.

6c). Results of calculations (Table II) demonstrate that, regardless of the oxidation state of the surface, adsorption of oxidative species on carbon atoms is significantly less favorable than over the area uncovered by LCCs. These results can be explained in another way. In the LCC, carbon atoms are $sp^1$ hybridized, and all sp-orbitals are directed along the chains. Thus, the overlap between the orbitals of adsorbed molecules is minimal. On the other hand, areas near the LCC always have out-of-plane orbitals suitable for the formation of robust bonds with oxidative species. In the case of pristine silver surfaces, there are 4d orbitals. In the case of passivation by oxygen, the orbitals have lone pairs of electrons, and in the case of passivation by hydroxyl groups, hydrogen bonds can form (see Fig. 6c). Therefore, in water or air environments, all oxidative species will preferably interact with the areas outside the LCC instead of oxidizing the carbon chains. In the case of a pristine silver surface, the LCC first oxidized the silver atoms, and all oxidative species will form robust noncovalent bonds with the oxidized surface (see Table II) instead of oxidizing the LCC.

**Conclusions**

The combined experimental and theoretical study of the chemical bonding and electronic structure of Ag and Ag@LCC nanoparticles prepared by pulse laser ablation (PLA) in water demonstrated the formation of stable monoatomic linear carbon chains on the surface of silver. XPS C 1s spectra showed that LCC atoms are not oxidized on Ag-nanoparticles. XPS Ag 3d and valence band spectra measurements led us to conclude that linear carbon chain atoms formed robust Ag-C bonds. The presence of LCC on the silver surface did not prevent oxidation of the uncovered metallic surface. Theoretical modeling demonstrated that the stability of LCC on silver was due to a strong preference for the physical adsorption of oxidative species uncovered by carbon than by adsorption on the carbon chains. This model demonstrates the possibility of protecting chemically unstable low-dimensional systems by forming composites that make the material much more attractive for oxidation and adsorption of oxidative species.

**Acknowledgements**

The XPS measurements were supported by the Russian Science Foundation (Project 14-22-00004).